\documentclass{optica-article}

\journal{opticajournal} 

\articletype{Research Article}

\usepackage{lineno}

\begin{document}

\title{Wide-Field Reflective Imaging with an Epi-Illumination Multi-Camera Array Microscope}

\author{Xiangjiang Bao,\authormark{1,2,*} Lucas Kreiss ,\authormark{1} Clare B. Cook,\authormark{1} Haoyu Gong,\authormark{1} Mark Harfouche,\authormark{2} and Roarke Horstmeyer,\authormark{1,2}}

\address{\authormark{1}Department of Biomedical Engineering, Duke University, Durham, North Carolina 27708, USA\\
\authormark{2}Ramona Optics Inc., 1000 W Main St., Durham, North Carolina 27701, USA\\}

\email{\authormark{*}xiangjiang.bao@ramonaoptics.com} 


\begin{abstract*} 
We present an epi-illumination multi-camera array microscope (epi-MCAM) designed for wide-field reflective imaging of non-transparent samples. The epi-MCAM contains 24 tightly packed and synchronized epi-illumination microscope units, arranged in a $4 \times 6$ planar array at 18 mm spacing. Each unit contains a unique CMOS image sensor (13 megapixels each), an objective and tube lens pair, and a beamsplitter and epi-illumination light path. An epi-MCAM capture cycle produces a stitched image covering $72 \times 108~\mathrm{mm}^2$ at a micrometer scale resolution down to 2.46~$\mu$m. To image samples exceeding this native field of view, we translate the entire array across the sample surface to enable high-resolution coverage of large objects. We demonstrate the system's ability to image both flat and three-dimensionally structured reflective samples, such as semiconductor wafers and printed circuit boards, which highlight the epi-MCAM's strong potential within industrial inspection applications.

\end{abstract*}

\section{Introduction}
High-resolution inspection of large-area samples, such as full semiconductor wafers or circuit boards, large tissue sections, and cultural heritage objects, remains a persistent challenge in optical microscopy \cite{1_huang_automated_2015,2_shankar_defect_2005}. It is challenging and time-consuming to image such large, macroscopic samples that can span across tens of centimeters at fine, microscopic resolution with existing imaging systems. Single acquisitions from conventional microscopes can only resolution sub-micrometer details across relatively small, millimeter-sized (or smaller) fields of view (FOV). Mechanical step-and-scan methods are thus required to cover the large imaged area, which for very large objects can take many minutes or hours to complete.

Achieving large-FOV imaging without compromising on resolution has long been a goal in the field of optical imaging systems \cite{3_zhang_simultaneous_2021,4_park_review_2021}. The trade-off between resolution and FOV is fundamentally constrained by the limited space-bandwidth product (SBP) \cite{5_lohmann_spacebandwidth_1996} of optical systems: as resolution increases, the imageable area decreases. Standard microscopes are typically limited to a total of 10--50 million resolvable spatial points (10--50 megapixels) \cite{4_park_review_2021}. 

Instead of using a single extremely large objective lens, which becomes prohibitively expensive and complex to correct for optical aberrations and to align, many systems have adopted a single high-resolution camera combined with mechanical scanning to overcome FOV limitations \cite{6_zheng_05_2014,7_korompili_portable_2018}. Scanning approaches require dense, step-by-step movements and stabilization at each position to ensure sufficient image coverage, leading to increased acquisition time and greater susceptibility to motion-induced artifacts.

Beyond mechanical scanning, several computational imaging strategies have been proposed to overcome SBP limitations, particularly in biological imaging. Many of these approaches rely on varying the illumination of the sample. For example, structured illumination microscopy (SIM) \cite{8_saxena_structured_2015,9_heintzmann_super-resolution_2017} enhances resolution by illuminating the sample with patterned light (typically sinusoidal), allowing the recovery of high spatial frequency content beyond the native resolution of the imaging system. However, SIM’s effective FOV remains constrained by the objective’s native imaging area and is typically limited to thin, fluorescent, and transparent samples. Fourier ptychographic microscopy (FPM) \cite{10_zheng_wide-field_2013,11_ou_quantitative_2013} is another method for increasing SBP, where the sample is illuminated under different oblique angles to synthetically expand the system's numerical aperture (NA). FPM supports high-resolution and quantitative phase imaging and has demonstrated potential for imaging reflective samples. However, similar to SIM, the FOV in FPM is still fundamentally limited by the native NA of the imaging optics. Moreover, fully realizing the potential of SIM and FPM systems often requires sophisticated reconstruction algorithms and stable, multi-frame data acquisition \cite{12_chen_superresolution_2023,13_lal_structured_2016,14_yeh_experimental_2015}, which can be computationally and experimentally demanding.

The use of multiple microscopes to image large areas in parallel has also been demonstrated in the recent past \cite{15_Harfouche:23,16_kim_rapid_2024,17_thomson_gigapixel_2022,18_Yang:25,19_yang_multi-modal_2023,20_kreiss_recording_2025}. Much of this prior work demonstrates how "multi-camera array microscopes" (MCAMs) provide a straightforward and efficient solution to expand the SBP of an optical imaging system, by using multiple synchronized imaging units to simultaneously capture different regions of a sample. Rather than relying on a single large-aperture system, the MCAM distributes the imaging task across a tiled array of compact and modular imaging units. This approach not only demonstrates a linearly scalable technology for large-area imaging, but also takes advantage of small optical components that are easier and less expensive to manufacture and integrate compared to their large counterparts. 

In most demonstrated MCAM systems to date, trans-illumination is used to capture images and video, typically of thin biological specimens. In a recent high-resolution configuration (0.3 - 0.5 numerical aperture)~\cite{16_kim_rapid_2024}, the imaging optics and sensor array are positioned above the sample and illumination is provided from below by an LED array. In lower resolution MCAM systems with larger working distances, LED boards are sometimes placed between the sample and MCAM's objective lenses to provide illumination from the same side as the imaging optics~\cite{21_zhou_parallelized_2023}. Unfortunately, such a layout introduces two key challenges: This configuration is difficult to accommodate high-NA lenses, which require short working distances; additionally, the oblique lighting geometry can cause non-uniform illumination across the array, resulting in inconsistencies in image brightness and quality among different imaging units. Another approach places rings of LEDs surrounding each imaging aperture directly beneath the lens plane or larger LED rings around the entire MCAM \cite{20_kreiss_recording_2025}. While this enables reflective illumination on highly diffuse objects, the annular lighting geometry often leads to brighter edges and darker centers in the captured images when the working distance is short, thereby reducing the uniformity and reliability of intensity-based measurements. Moreover, this type of external illumination is a poor choice for fluorescent excitation, which ideally is focused evenly across the FOV of each camera.

Introducing epi-illumination \cite{22_ishmukhametov_simple_2016,23_edwards_epi-illumination_2014} offers an effective solution to many of the above imaging challenges. In an epi-illumination configuration in conventional microscopes, illumination light passes through the same objective lens that is used for imaging, thereby illuminating the sample from the same side as where imaging occurs. In other words, the optical paths for illumination and imaging are coaxially aligned. Integration of the light source into the optical path ensures relatively uniform illumination of thick, non-transparent and reflective samples via the expansive effect of the objective lens across its associated FOV. It also enables a more compact optical design. Importantly, a coaxial ``epi" arrangement eliminates the need to fit bulky illumination components between the sample and the objective lens. The reduced working distance, in turn, permits the use of larger NAs for imaging thick or reflective objects, which is critical for enhancing spatial resolution and image quality \cite{24_murphy_fundamentals_2012}. In short, epi-illumination not only improves illumination uniformity, but also enables the high-NA optics necessary for high-resolution reflective imaging. 

However, implementing epi-illumination in a dense multi-camera array presents significant challenges. Unlike external illumination, epi-illumination requires that excitation light be delivered internally through each objective. While achieving an epi-illumination design can be relatively straightforward for standard microscopes with direct access to the imaging path behind the objective lens, tightly packing dozens of objectives into a compact array leaves very little physical space for routing illumination components such as beamsplitters, prisms, or light guides. In particular, for imaging units located in the center of the MCAM array, there is no direct side access for introducing illumination, making it difficult to couple light into these objectives without interfering with adjacent cameras. 

In this work, we present a novel solution to the above challenges to produce a new format of ``epi-MCAM" - a compact array of objective lenses and illumination units that can rapidly image large non-transparent objects at high resolution and with high uniformity. Apart from achieving epi-illumination within a dense imaging array, the epi-MCAM also employs a new objective lens design based on an infinite-conjugated pair of relatively high-NA lenses, which greatly increases the magnification compared to many previous non-scanning MCAM configurations with single lenses (typically $M \approx 0.2$) \cite{15_Harfouche:23}. To demonstrate the epi-MCAM, we rapidly imaged a full 300 mm diameter semiconductor wafer, producing a stitched composite image comprising 75 gigapixels (GP). To demonstrate thick sample image capture, we used the epi-MCAM to 3D scan a complex NVIDIA GPU board with components of varying thicknesses over a 72 $\times$ 108 mm$^2$ FOV by acquiring multiple focal planes and digitally merging the sharpest regions to ensure complete depth coverage. The demonstrated system operates at a measured magnification of 1$\times$ and achieves an effective lateral two-point optical resolution of approximately 2.46 $\mu$m, with direct room to improve to higher resolutions and achieve integrated fluorescence excitation in updated future designs.

\section{Epi-illumination design for MCAM}

Our primary motivation for implementing epi-illumination in an MCAM is to enable parallelized reflective and fluorescence excitation imaging using high-NA lenses. The NA of an optical system is defined as 
$\mathrm{NA} = n \cdot \sin(\theta)$, where $\theta$ is the maximum half-angle of the light cone that the objective can collect from a point on the sample~\cite{24_murphy_fundamentals_2012}. 
To increase the angle $\theta$ for a clear aperture with a fixed diameter (specified by the MCAM's inter-camera spacing), the working distance must be reduced. The short working distance that is created when operating the MCAM at high resolutions (micrometer to (sub-micrometer) limits the physical space between the lens and the sample, which in turn makes it difficult to incorporate additional illumination hardware and poses a key challenge for reflective MCAM imaging. 

Due to the gap between adjacent microscopes in the array, it is clear that a lens with a magnification greater than 1 will also result in a gap between adjacent FOVs~\cite{15_Harfouche:23}. 
Thus, when operating the MCAM with relatively large magnification and relatively high resolution, a small amount of lateral scanning is needed to fully image large specimens. In our implementation, each imaging unit uses a pair of identical lenses (Edmund 58-207, $f = 25$~mm, $f/\# = 2.5$) arranged in an infinite-conjugate setup, resulting in a system magnification of $M = 1$. Based on the pixel-limited resolution defined by the full-pitch criterion 
$r = 2\delta / M$, where $\delta$ is the pixel width, a system using CMOS sensors with pixel sizes between 0.7~$\mu$m and 1.0~$\mu$m 
can achieve a full-pitch optical resolution of approximately 1.4~$\mu$m to 2~$\mu$m, which is comparable to that of standard $4\times$ to $10\times$ microscope objectives 
with numerical apertures in the 0.15 to 0.3 range~\cite{15_Harfouche:23}. In this work, we employ CMOS sensors with $\delta = 1.1~\mu$m pixels, which sets our maxmimum achievable full-pith optical resolution as 2.2~$\mu$m.

\begin{figure}[htbp]
\centering\includegraphics[width=12.5cm]{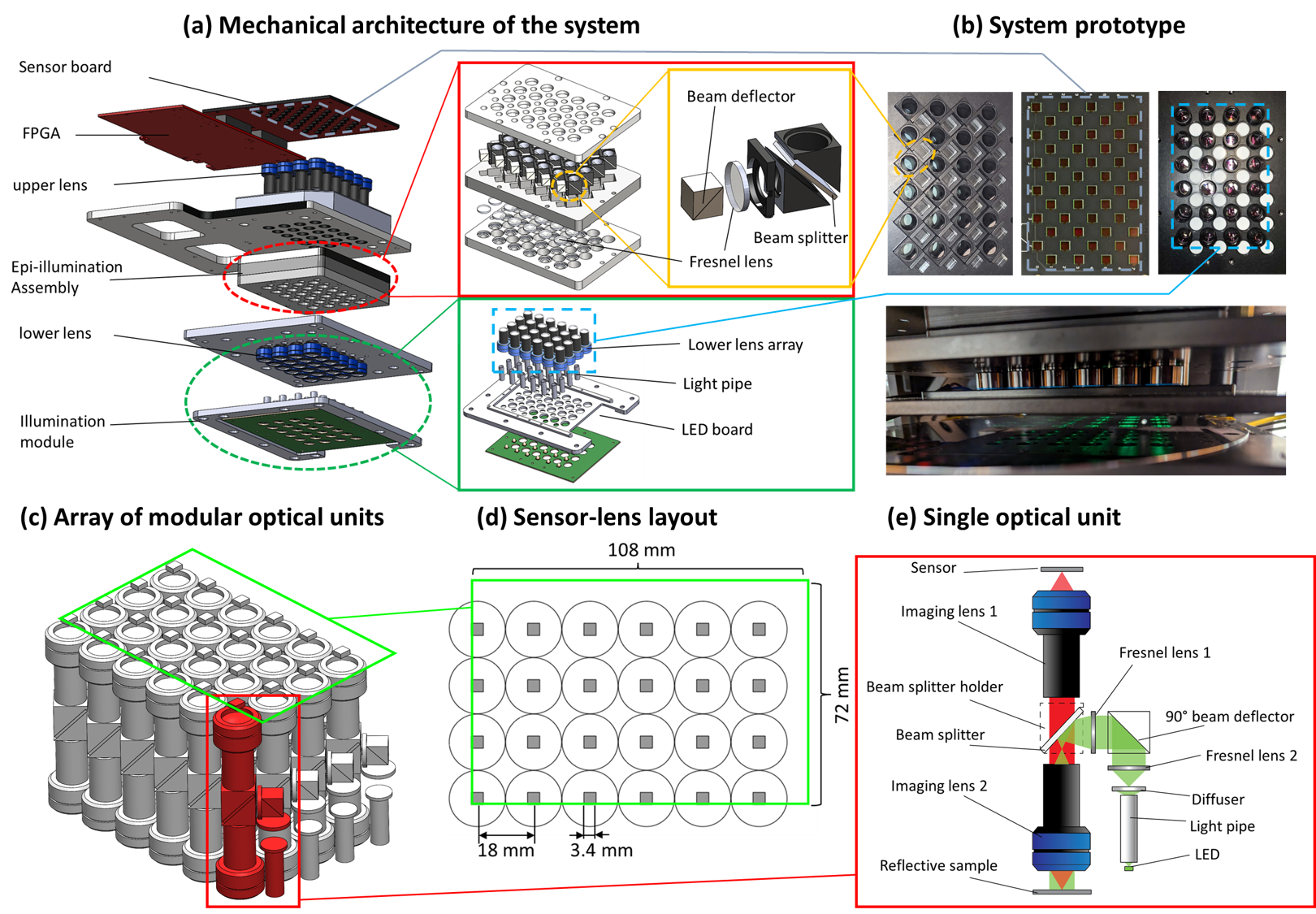}
\caption{Mechanical and optical design of the epi-illumination multi-camera array microscope (epi-MCAM). 
(a) Exploded view of the full epi-MCAM assembly. 
(b) Photograph of the epi-MCAM prototype, including views of selected components.
(c) 3D diagram of the $4 \times 6$ optical unit array in the epi-MCAM, each comprising an imaging lens pair and epi-illumination optics (see panel e). 
(d) Top view of the sensor-lens layout. An image with a $72 \times 108$~mm$^2$ field of view (FOV) is acquired by scanning to fill the gaps between adjacent sub-FOVs. 
(e) Optical layout of a single optical unit. A 50:50 beam splitter is used to combine the illumination and imaging light paths.}
\label{fig1}
\end{figure}

The overall architecture of our epi-MCAM system is shown in Fig.~\ref{fig1}(a). Our first prototype utilizes a $6 \times 8$ array of CMOS sensors (Onsemi AR1335, $4208 \times 3120$ pixels, 
$\delta = 1.1~\mu$m) for synchronized data acquisition. These sensors are arranged in a compact grid on a single circuit board. In this work, only half of the sensors are activated 
since the optical array comprises only $4 \times 6$ units. A field-programmable gate array (FPGA) is mounted beneath the sensor array for high-speed data transmission. 
Since the FPGA supports data transfer at up to approximately 5~GP/s (or 5~GB/s)~\cite{16_kim_rapid_2024,21_zhou_parallelized_2023}, up to 8 synchronized frames from all 48 sensors can be captured per second. 
The frame rate can be further increased by applying pixel binning or reducing each sensor’s active area. In our work, all measurements are performed in high-resolution mode, 
where each active sensor captures a square region of $3072 \times 3072$ pixels.

This work's prototype epi-MCAM design has a center-to-center inter-camera spacing of 18~mm. This inter-camera distance specifies the maximum allowable lens diameter. To achieve epi-illumination within this limited spacing, we employ multiple custom plate holders, each with optimized mounting holes for the optical components (Fig.~\ref{fig1}(a)). 
Illumination is provided by a dedicated LED board populated with a $4 \times 6$ array of high-power LEDs. Each LED contains three independently controlled spectral channels centered at 430~nm, 530~nm, and 580~nm, each with a bandwidth of $\sim$30~nm and driven with up to 3~W of electrical input. Somewhat counter intuitively, the LED array is positioned such that light shines upward, away from the sample, and towards the infinity space between the objective and tube lens for each unit microscope.

The optical path of each individual optical unit in the array is illustrated in Fig.~\ref{fig1}(e). 
An inner-polished light pipe is used to guide the light from the LED into the illumination path. 
By joining two identical right-angle prisms with an edge length of 8~mm into a cube, we construct a 90$^\circ$ beam deflector, where the reflective surface is formed by coating aluminum on the slanted face of one prism. 
Although only the lower prism actively participates in beam reflection, using a symmetric cube configuration improves mounting precision and mechanical stability within the plate holder. 
Two Fresnel lenses (Thorlabs FRP0510) are employed to collimate the illumination and generate a uniform, high-intensity light pattern on the sample. 
To achieve epi-illumination, a 50:50 beam splitter reflects half of the LED light toward the sample and transmits half of the light returning from the sample to the imaging sensor. 
Additionally, the beam splitter holder is fabricated from aluminum with a black-anodized surface and further coated with a light-absorbing material to minimize stray reflections.

The system is mounted on a 3-axis motorized Zaber stage platform for image acquisition. 
The scanning platform includes two parallel motorized linear stages for motion along the x-axis (LRT0500HL-E08CT3A, 500~mm travel range, 0.39~$\mu$m resolution), 
and one motorized stage (LRT0750HL-E08CT3A, 750~mm travel range, 0.39~$\mu$m resolution) for y-axis translation. 
For z-axis scanning, a high-precision linear stage (LRT0100AL-E08CT3A, 100~mm travel range, 0.12~$\mu$m resolution) is used to ensure accurate focus control. 
Motion along all three axes is coordinated through a multi-axis universal motor controller (X-MCC4), allowing synchronized and automated positioning during scanning. 

Fig.~\ref{fig1}(d) shows the layout of the sensors and lenses in our current epi-MCAM. 
The circles indicate the outer diameter of each imaging lens, while the gray square blocks represent the sub-FOVs, corresponding to the active sensor regions since the magnification is $M=1$ for each unit. 
To fill the gaps between sub-FOVs, an $8 \times 8$ scanning sequence is performed, where each snapshot overlaps with its neighbors by $\sim$25\% to facilitate accurate stitching. 
The full FOV of $72 \times 108$~mm$^2$ (indicated by the green block) is reconstructed by stitching the acquired image data into final composite frames.

\section{Imaging workflow}

\begin{figure}[htbp]
\centering\includegraphics[width=12cm]{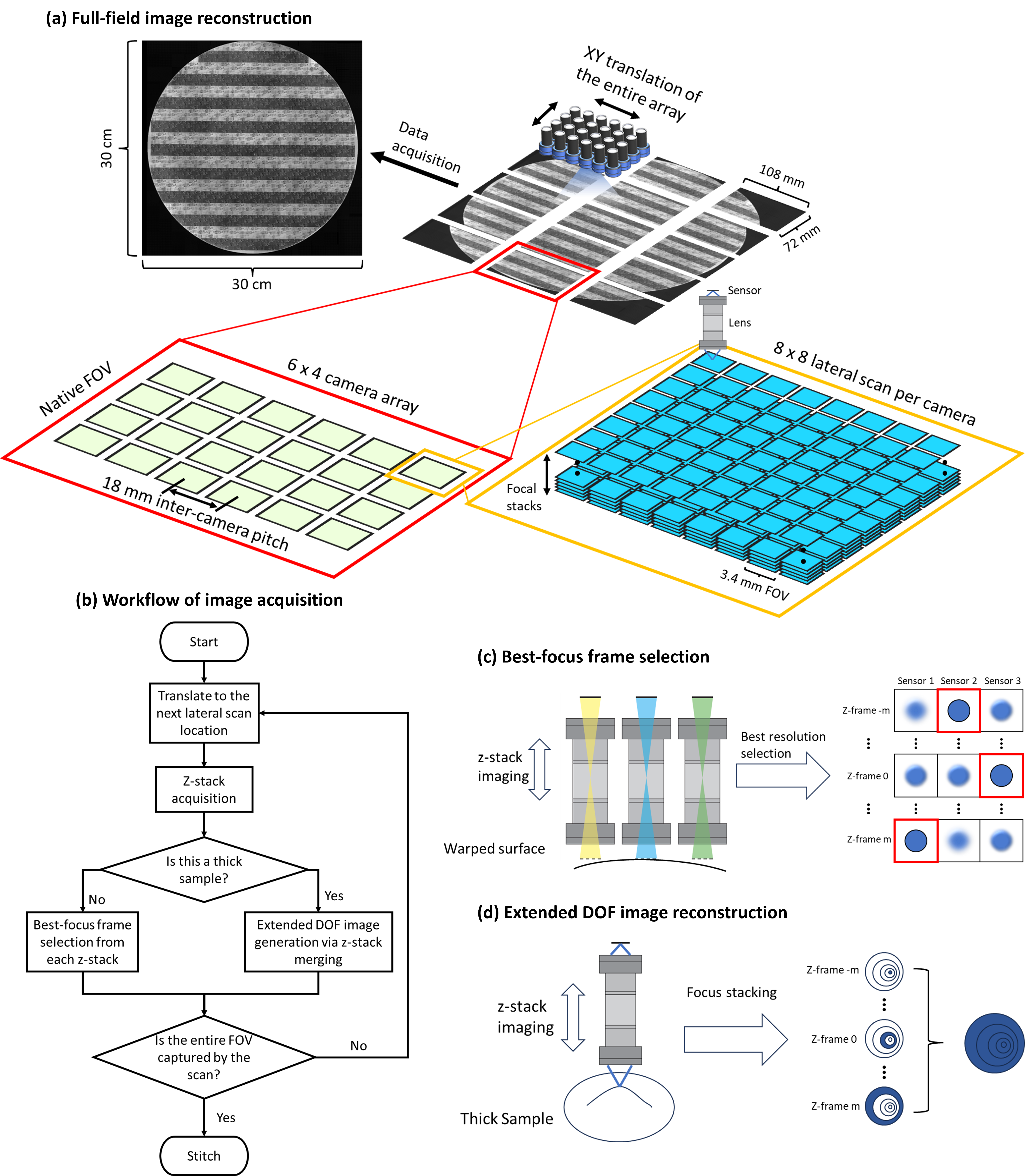}
\caption{Diagrams of epi-MCAM image acquisition. 
(a) Illustration of the data acquisition process for reconstructing an image that exceeds the native FOV (108$\times$72~$\mathrm{mm}^2$) of a single imaging cycle. The sample shown here covers an FOV of 30$\times$30~$\mathrm{cm}^2$.
(b) Description of the scanning workflow for acquiring a full-field image for both flat and thick samples.
(c) Diagram of selecting the best-focused image from each z-stack for flat sample imaging with a wrapped surface. The blue regions indicate the in-focus areas. 
(d) Diagram of merging images in a z-stack to generate an extended depth-of-field (eDOF) image for thick sample imaging.}
\label{fig2}
\end{figure}

As shown in Fig.~\ref{fig2}(a), the system uses rapid z-stacking at each scan position to acquire in-focus image data across macroscopic samples that are not perfectly flat.
We use an imaging numerical aperture of $\mathrm{NA}=0.2$. 
The associated depth of field (DOF) of each microscope unit is thus 
$\lambda / \mathrm{NA}^2 = 13.2~\mu\mathrm{m}$ at $\lambda = 530~\mathrm{nm}$, 
which is relatively shallow (see Fig.~3(c)). To address this limitation, z-stack imaging is employed to capture information across multiple focal planes. Our system employs infinite-conjugate imaging pairs, ensuring that the magnification of each imaging unit remains nearly constant despite focal plane offsets. 
As a result, z-stack acquisition offers a simple yet robust approach to capturing in-focus data across all units. 

We propose a workflow for full-field image acquisition that is applicable to both relatively flat 2D surfaces and thicker 3D samples (Fig.~\ref{fig2}(b)). For relatively flat 2D surfaces, software was developed to rapidly select the best-focused image from each acquired z-stack per each of the 24 microscope units per position. To achieve this, an image-based sharpness metric is computed per z-stack image and the maximum value is selected, as shown in Fig.~\ref{fig2}(c). For thicker samples, software was also developed to perform extended depth-of-field (eDOF) reconstruction from each z-stack acquisition, as shown in Fig.~\ref{fig2}(d). 

The $8 \times 8$ resulting image tiles (either of best-focus or eDOF) from each individual microscope unit are first stitched to form 24 composite images, each corresponding to the scanning area of one imaging unit. 
These 24 images are then further stitched to generate the complete full-field image. If the sample exceeds the native FOV of a single imaging cycle (72 $\times$ 108 mm$^2$), the entire system is repositioned, and the imaging workflow is repeated until the entire sample is covered (Fig.~\ref{fig2}(a)). Otherwise, the translation step of the entire array to acquire native FOVs from different regions is skipped.

In practice, the system could be calibrated once by recording the best focal plane position for each camera, so that subsequent data collection can be restricted to those planes, thereby reducing both acquisition time and processing load. 
Such calibration would effectively address the inherent focal plane offsets among the imaging units, but it cannot fully compensate for variations introduced by surface wrapping of the sample.

\section{Results}

To evaluate the performance of epi-illumination imaging within an epi-MCAM, we captured various calibration, assessment and experimental data with our $4 \times 6$ prototype array. Image data from the 24 individual camera units were synchronously routed to an FPGA with less than a 6 micro-second inter-camera offset. The FPGA is connected to a Thunderbolt-to-PCIe expansion chassis. This chassis is also connected by a Thunderbolt cable to a laptop equipped with 125~GiB of RAM and 4~TB of storage. The high-volume data generated by the sensors (approximately 0.21~GP per frame) are transmitted to the laptop via the PCIe and Thunderbolt interface for storage and processing. 
When operated with lateral scanning, movement resumes only after all data from the current location have been successfully written.

The FPGA provides programmable control over all sensor parameters, including exposure time and digital gain, through an integrated graphical user interface (GUI) on the laptop. The same GUI also provides programmable control of the 3-axis Zaber stage, enabling flexible scanning configurations. We employed the high-performance Huygens software from Scientific Volume Imaging~\cite{26_noauthor_homepage_nodate} to generate all stitched images. Its integrated automatic vignetting correction algorithm substantially improved the continuity and overall quality of the stitching results.

\subsection{Optical validation}
\begin{figure}[htbp]
\centering\includegraphics[width=12.5cm]{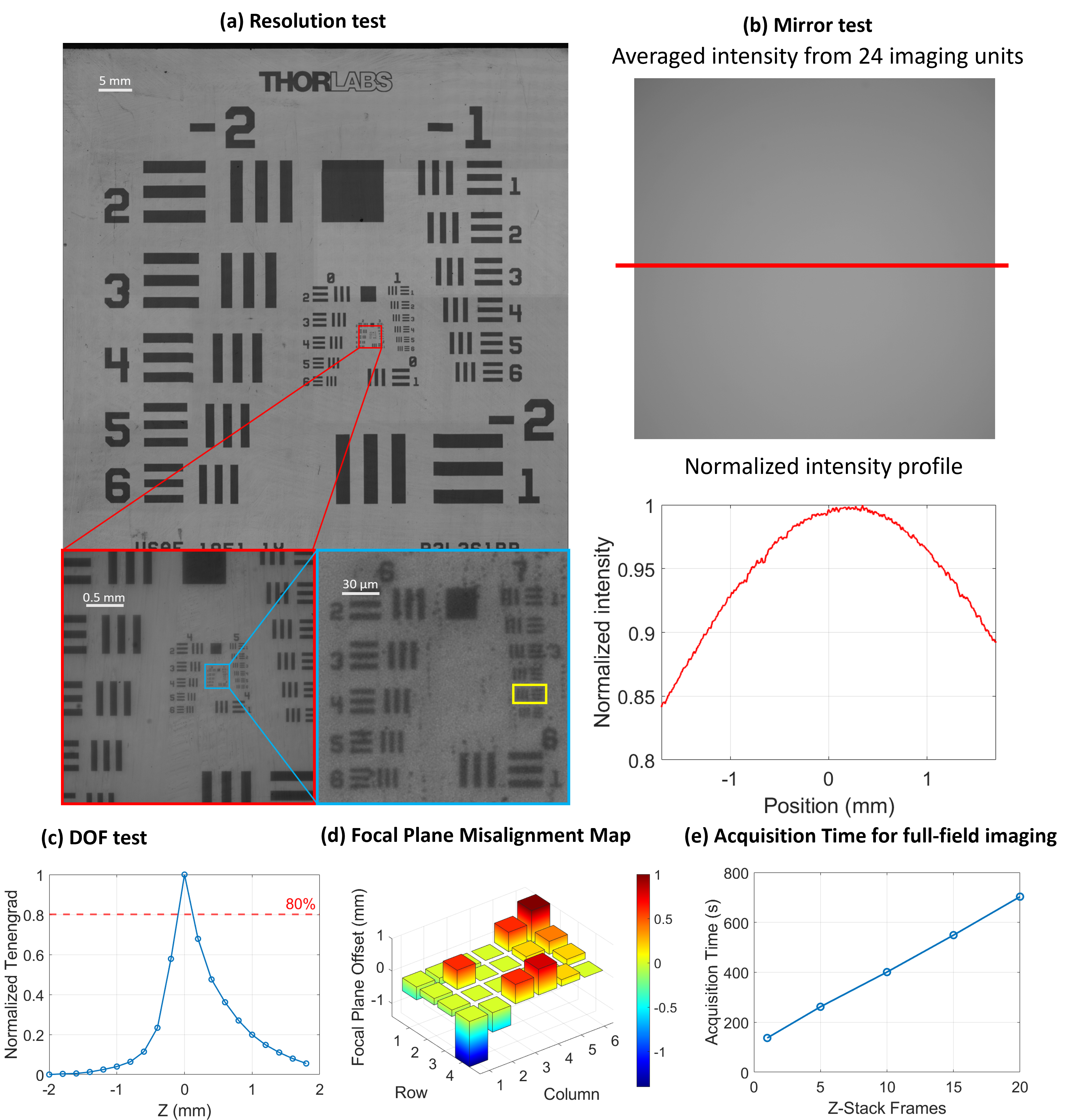}
\caption{Imaging performance characterization. 
(a) Image of a 1951 USAF resolution target acquired with our system. The image is stitched from the central $4 \times 4$ imaging units. The inset within the blue box is displayed with a 40\% enhancement in contrast. The yellow block highlights the region containing the smallest resolvable element with a line width of 2.46~$\mu$m. 
(b) Averaged intensity from 24 imaging units capturing a mirror. The normalized central profile is shown as the red solid line. 
(c) Evaluation of the resolution degradation of a single imaging unit as a function of the z-position using the Tenengrad sharpness metric. The red dashed line indicates 80\% of the maximum sharpness value. 
(d) Map of focal plane offsets for each imaging unit relative to the central unit (row~2, column~3). 
(e) Acquisition time required to capture a full-field ($72 \times 108$~mm$^2$) image using different numbers of z-stack frames.}
\label{fig3}
\end{figure}

To quantitatively verify the performance specifications of our constructed epi-MCAM system, we conducted a series of calibration experiments, as summarized in Fig.~\ref{fig3}. 
To evaluate the resolution, we imaged a 1951 USAF resolution target (Fig.~\ref{fig3}(a)) following the flat sample imaging workflow in Fig.~\ref{fig2}(b). 
The system performed an $8 \times 8$ scanning sequence, acquiring 20 z-stack frames for each sub-image with a 0.2~mm axial step size. 
The yellow block in Fig.~\ref{fig3}(a) highlights the region containing the smallest resolvable element with a line width of 2.46~$\mu$m, which approximately matches the theoretical full-pitch resolution limit set by the pixel size (2.2~$\mu$m). 
For stitching, the frame with the highest resolution was selected from each z-stack using the Tenengrad sharpness metric~\cite{27_her_research_2019}:

\begin{equation}
T = \sum_{x,y} \left( G_{x}(x,y)^{2} + G_{y}(x,y)^{2} \right),
\end{equation}

where $G_{x} = S_{x} * I$ and $G_{y} = S_{y} * I$, 
with $S_{x}$ and $S_{y}$ denoting the Sobel operators in the horizontal and vertical directions, respectively. $I$ denotes the original image, and $*$ represents convolution.

\begin{equation}
S_{x} =
\begin{bmatrix}
-1 & 0 & 1 \\
-2 & 0 & 2 \\
-1 & 0 & 1
\end{bmatrix},
\quad
S_{y} =
\begin{bmatrix}
-1 & -2 & -1 \\
0 & 0 & 0 \\
1 & 2 & 1
\end{bmatrix}
\end{equation}

To evaluate the uniformity of the intensity across the FOV of each imaging unit, the USAF target was replaced with a mirror. The average imaged intensity response across all 24 units is shown in Fig.~\ref{fig3}(b)). 
The central normalized intensity trace profile shows that the average intensity decreased by $\sim$15\% at the edge relative to the center, indicating that the imaging units maintain relatively high uniformity across each full per-camera FOV.

We additionally measured the resolution degradation as a function of the z-position. In the test shown in Fig.~\ref{fig3}(c), we imaged Groups~2 and~3 of the 1951 USAF resolution target and captured a 20-frame z-stack with a 0.2~mm axial step size, which was identical to the acquisition settings used for the resolution evaluation. The depth range over which the normalized Tenengrad sharpness metric remained above 80\% of its maximum was approximately 0.24~mm. 

To quantify the focal plane misalignment across the imaging units, we captured a 20-frame z-stack image sequence of a large semiconductor wafer using the same axial step size. 
The wafer was sufficiently large to span all imaging units within the array. 
For each unit, the sharpest image within the z-stack was identified based on the Tenengrad metric, and the corresponding frame index was recorded to estimate the relative focal plane offsets. 
The results of our focal plane misalignment measurement are demonstrated in Fig.~\ref{fig3}(d). 
The largest deviations are observed in the diagonal corner units (row~4, column~1 and row~1, column~6), which is likely caused by slight bending of the metal plate used to mount the epi-MCAM system onto the scanning stage. 
The maximum focal plane offset measured across the array is 2.4~mm, which explains the necessity of z-stack acquisition even for flat samples. 

Fig.~\ref{fig3}(e) plots the total system acquisition time for a continuous area (108$\times$72~$\mathrm{mm}^2$, $\sim$6.4~GP per stitched 2D frame) as a function of the number of z-stack frames captured across the entire array. 
Capturing a full-field image with 20 z-stack frames requires 703~s, whereas a single-frame acquisition (no z-stack) only takes 137~s. 
These results demonstrate the system’s potential to significantly reduce imaging time by minimizing the required number of z-stack frames, which is an improvement that can be achieved by properly calibrating the focal plane of each imaging unit.

\subsection{Experimental Imaging of a 300 mm diameter Semiconductor Wafer}

\begin{figure}[htbp]
\centering\includegraphics[width=12.5cm]{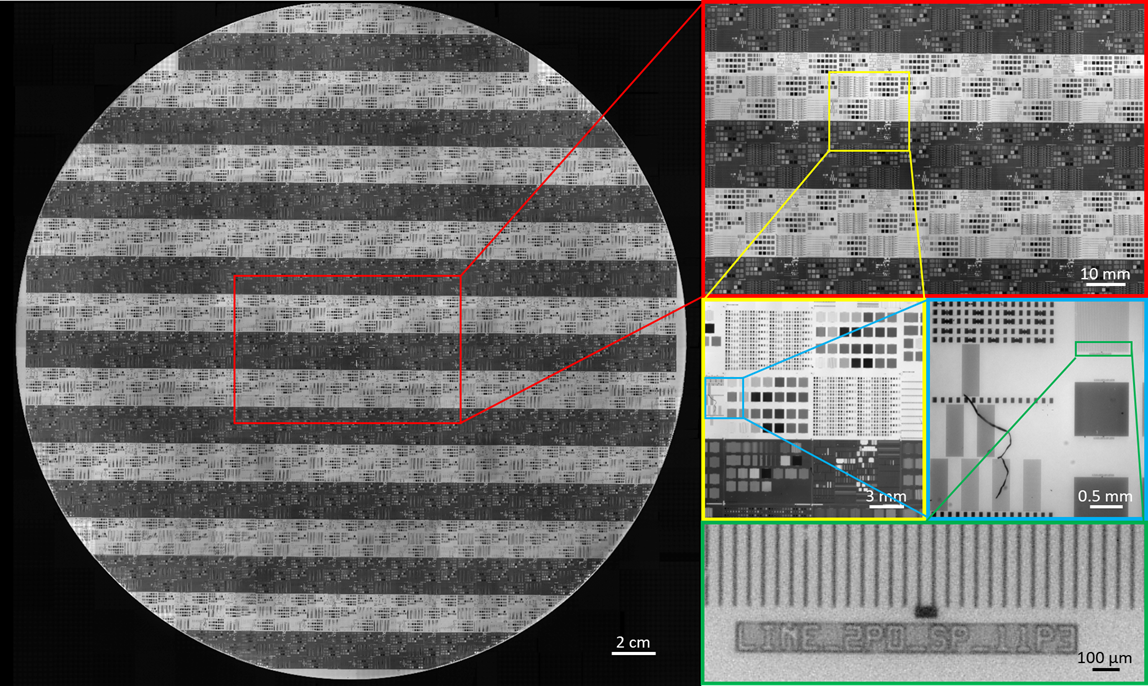}
\caption{Stitched image of a 300~mm-diameter semiconductor wafer acquired using epi-MCAM. 
The image was stitched from $3 \times 5$ tiled full-field FOVs, corresponding to a total of $\sim$75~gigapixels.}
\label{fig4}
\end{figure}

We captured and stitched the image of a 300~mm-diameter semiconductor wafer to demonstrate the system’s capability in large flat sample imaging (Fig.~\ref{fig4}). 
The entire imaging process followed the flat sample imaging workflow described in Fig.~\ref{fig2}(b). 
A total of $3 \times 5$ full-field images were acquired to cover the entire wafer, following the data acquisition process shown in Fig.~\ref{fig2}(a). 
For each full-field image, an $8 \times 8$ scanning sequence was performed. The \(8\times8\) scan pattern ensures approximately 25\% overlap between adjacent sub-images, which facilitates a reliable stitching process. 
The total acquisition time can be reduced by decreasing the number of scans to a \(6\times6\) sequence.
At each lateral scan position, a 20-frame z-stack was acquired with a 0.2~mm axial step size. 
The sharpest image within each z-stack was selected based on the Tenengrad metric for further processing. 
The figure presents magnified views at multiple scales, with the lines in the bottom green block measuring approximately 10~$\mu$m in width. The total number of pixels acquired for all focal stacks reached approximately \(4050~\mathrm{GP}\). After best-focus selection, the total number of best-focus images was reduced to \(\sim203~\mathrm{GP}\). After stitching, the final reconstructed image contains \(\sim75~\mathrm{GP}\) pixels.

\subsection{Experimental Imaging of a Thick PCB Board}

\begin{figure}[htbp]
\centering\includegraphics[width=11cm]{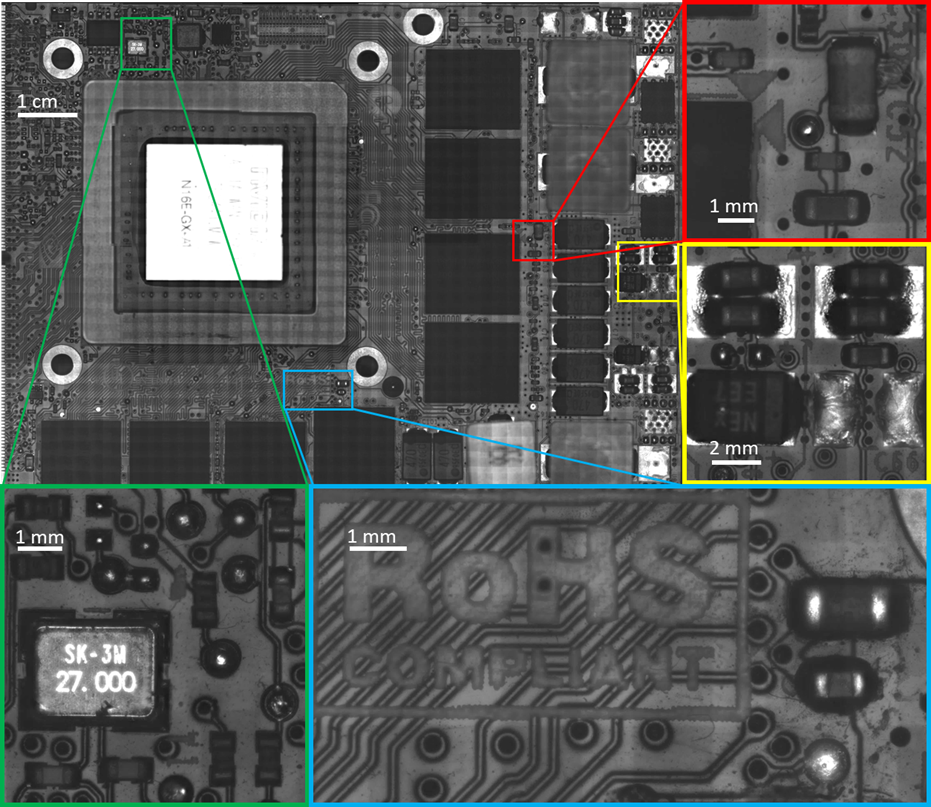}
\caption{Stitched image of an NVIDIA GPU circuit board acquired using epi-MCAM. 
The board includes electronic components with varying heights, and the stitched image contains a total of $\sim$6.4~gigapixels.}
\label{fig5}
\end{figure}

We used the same scanning parameters as in the wafer imaging to acquire a full-field image of an NVIDIA GPU circuit board, which contains various electronic components with different heights (Fig.~\ref{fig5}). 
The data acquisition involved an $8 \times 8$ lateral scan with a 20-frame z-stack acquired at 0.2~mm axial increments. 
Following the thick-sample imaging workflow described in Fig.~\ref{fig2}(d), we generated an all-in-focus image by applying the PMax algorithm from Zerene Stacker~\cite{28_noauthor_home_nodate}. 
This algorithm selects the pixel with the highest local contrast from each focal plane to produce an eDOF image. 

The magnified subfigures in Fig.~\ref{fig5} demonstrate features at different heights across the board. 
For example, the green box shows a crystal oscillator approximately 1~mm in height. 
In the blue box, capacitors of two different heights (0.8~mm and 0.5~mm) are clearly resolved, along with fine surface features on the PCB such as the RoHS compliance symbol. 
Given the measured DOF of 0.24~mm based on the Tenengrad metric, the effective DOF has been significantly extended through focus stacking. 
However, the central GPU chip appears overexposed and slightly out of focus due to its mirror-like surface and relatively large height, which measures approximately 3.6~mm. The total number of pixels acquired for all focal stacks reached approximately \(270~\mathrm{GP}\). After eDOF reconstruction, the total number of reconstructed images was reduced to \(\sim13.5~\mathrm{GP}\). After stitching, the final reconstructed image contains \(\sim6.4~\mathrm{GP}\) pixels.

\section{Discussion and conclusion}

In this work, we demonstrate a new format of large-area high-resolution imaging via a compact array of epi-illumination microscopes. We show how our newly designed epi-MCAM system can rapidly image both flat and thick samples across tens-of-centimeters scales, with over 30 cm demonstrated with a full reflective wafer, while maintaining micron-level resolution. The system demonstrates strong potential for imaging non-transparent samples at high resolution, which may be particularly interesting for example for semiconductor surface inspection. Moreover, the implementation of epi-illumination makes it possible for MCAM technology to support epi-fluorescence imaging~\cite{29_webb2012epi,30_kim_deep-brain_2017} and other techniques that rely on coaxial illumination, such as differential interference contrast (DIC) microscopy~\cite{31_pszonka_quantitative_2018,32_kandel_epi-illumination_2019}. In future updates. These extensions may also offer potential benefits for biological imaging.  

The primary difficulty that we faced in this work was rapidly processing the large volumes of acquired image data. Our current imaging protocol for flat samples requires the acquisition of deep z-stacks to ensure sharp focus across the entire array. Each z-stack typically contains 20 frames per position. This requirement jointly stemmed from the geometry of the sample as well as focal plane offsets between individual imaging units. A precise hardware calibration that brings all imaging units into a common focal plane could serve as an impactful short-term optimization of the epi-MCAM imaging hardware. For flat samples without surface wrapping, such as wafers, successful calibration can substantially reduce the necessary z-stack depth, lowering stack acquisitions from 20 frames to just 2--3 frames, which is adequate to compensate for an overall sample tilt of approximately 0.4~mm. As shown in Fig.~\ref{fig3}(e), the acquisition time scales linearly with the number of frames in each z-stack. A reduction in the number of frames would not only decrease the data storage requirements, but also significantly reduce the time needed for data transmission. For thick samples, integrating an autofocus capability into each individual lens could enable the system to capture surface information from curved samples while maintaining high focus quality, as we already discussed previously for other MCAM configurations~\cite{18_Yang:25,20_kreiss_recording_2025}. Additionally, techniques such as incorporating a designated phase mask into the optical path have the potential to extend the system’s DOF~\cite{33_yang_optimized_2007,34_castro_asymmetric_2004}, thereby reducing the number of frames required in each z-stack.  

Currently, the speed of data transmission also becomes a limiting factor for overall epi-MCAM scan speed, since our current system design does not proceed in scan step until all data from the current position is fully written to computer memory. In our implementation, the data transmission speed is 2.5~GB/s---approximately half of the maximum rate reported in Refs.~\cite{15_Harfouche:23,16_kim_rapid_2024}. This reduction is primarily due to our choice of transmission interface. Instead of using a direct PCIe connection from the FPGA, we routed the PCIe output through a Thunderbolt converter and transmitted the data via Thunderbolt. This approach was intended to overcome the physical constraints associated with cable movement during scanning. The data acquisition speed can be further improved through hardware upgrades in the data transmission strategy, such as replacing the Thunderbolt interface with a direct PCIe connection or employing parallel high-speed interfaces to simultaneously stream data from multiple FPGA channels.  

Another way to improve acquisition time and reduce data storage without modifying the hardware is to adjust the number of positions in the lateral scanning sequence. In this work, we performed an $8 \times 8$ scan (64 snapshots) for each imaging cycle. Each sub-image overlaps with its neighbors by approximately 25\% to ensure accurate stitching. The data acquisition speed can be further increased by reducing the number of scanning positions to $6 \times 6$ (36 snapshots), at the cost of lower coverage. This brings us to another discussion in data analysis: is image stitching truly necessary? While stitching all sub-images into a single large composite image is visually compelling, it is also time-consuming and inefficient in terms of both computational cost and memory usage. A useful analogy can be drawn from the YOLO network~\cite{35_hussain_yolo-v1_2023,36_jiang_review_2022,37_liu_object_2018}, which divides a large image into a grid, where each cell is responsible for predicting the presence of an object, along with its bounding box and class probabilities. This concept aligns with the data acquisition strategy of our MCAM system, which constructs a large image from grids of sub-images. In practice, individual sub-images can be used directly for training and detection, as the required information is already contained within them. This approach eliminates the need for image stitching in many deep learning-based applications, such as semiconductor surface inspection~\cite{38_chien_inspection_2020} and animal motion tracking~\cite{17_thomson_gigapixel_2022}.

\begin{backmatter}
\bmsection{Funding}
Research reported in this publication was supported by the National Institute of Mental Health (NIMH) of the National Institutes of Health under Award Number R43MH133521, the Office of Research Infrastructure Programs (ORIP), Office Of The Director, National Institutes Of Health of the National Institutes Of Health and the National Institute Of Environmental Health Sciences (NIEHS) of the National Institutes of Health under Award Number R44OD024879, the National Cancer Institute (NCI) of the National Institutes of Health under Award Numbers R44CA285197 and R44CA250877, a Duke-Coulter Translational Research Grant Award, and a National Science Foundation Award (2036439)

\bmsection{Acknowledgment}
The authors thank Shenzhen Yingfeng Optoelectronics and Shanghai Think Automation for helpful discussions regarding the fabrication.

\bmsection{Disclosures}
X.B. was a student at Duke University during this work and is currently employed by Ramona Optics, Inc.  
R.H. is both a faculty member at Duke University and a co-founder of Ramona Optics, Inc.  
M.H. is a co-founder of Ramona Optics, Inc.  
The remaining authors declare no conflicts of interest.

\bmsection{Data Availability Statement}
Image data presented in this paper are not publicly available due to their large volume, but can be obtained from the corresponding author upon reasonable request.

\end{backmatter}

\bibliography{epi_mcam}

\end{document}